\documentclass{vgtc}                          

\graphicspath{{figures/}{pictures/}{images/}{./}} 

\usepackage{times}                     

\usepackage{tabu}                      
\usepackage{booktabs}                  
\usepackage{lipsum}                    
\usepackage{mwe}                       

\usepackage{mathptmx}                  

\usepackage[table]{xcolor}
\definecolor{lightorange}{RGB}{255,230,200}   
\definecolor{medorange}{RGB}{255,210,160}    
\definecolor{deeporange}{RGB}{255,190,120} 

\onlineid{C-2178}

\vgtccategory{Research}

\vgtcinsertpkg
\usepackage{orcidlink}

\title{FlexiCamAR: Enhancing Everyday Camera Interactions on AR Glasses with a Flexible Additional Viewpoint}

\author{Ziming Li~\orcidlink{0009-0004-7529-7176} %
\and Hongji Li~\orcidlink{0000-0001-8638-7055} %
\and Jialin Wang~\orcidlink{0000-0002-1990-1293} %
\and Pan Hui~\orcidlink{0000-0001-6026-1083} %
\and Hai-Ning Liang~\orcidlink{0000-0003-3600-8955}\thanks{Corresponding author (hainingliang@hkust-gz.edu.cn)}}
\affiliation{\scriptsize Computational Media and Arts Thrust \\ The Hong Kong University of Science and Technology (Guangzhou), Guangzhou, China}

\teaser{
  \centering
  \includegraphics[width=\linewidth]{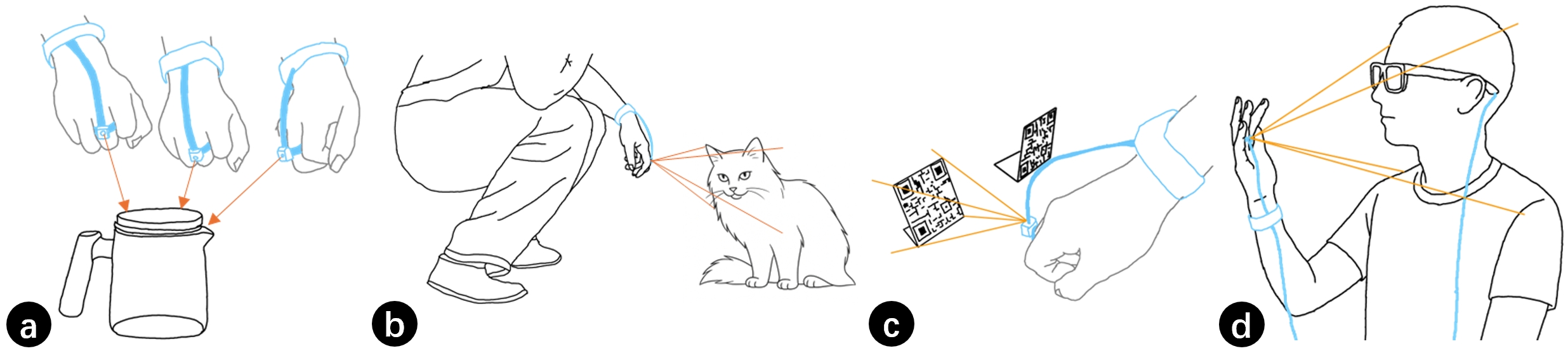}
  \caption{FlexiCamAR enhances everyday camera interaction on AR glasses by providing a flexible secondary viewpoint. This enables tasks such as: (a) effortless multi-angle photography of objects; (b) convenient low-angle captures; (c) efficient and targeted QR codes scanning; and (d) selfie photography or videoconferencing.}
  \label{fig:teaser}
}

\abstract{
    The recent emergence and popularity of consumer-grade augmented reality (AR) glasses from major technology companies highlight their potential to become the next daily computing platform. A dominant design trend in this context is the integration of a front-facing camera to deliver a first-person perspective. While this approach is intuitive, there is limited evidence that it is optimal (or sufficient) for supporting users in daily tasks. This paper explores a more effective camera interaction technique for AR glasses, which we term ``FlexiCamAR." This novel method aims to enhance both efficiency and the range of applications for AR glasses by offering flexible and comfortable secondary camera viewpoints. To investigate the applicability and usability of this approach, we developed a ring camera prototype that can be attached to users’ fingers. We then conducted a user study with 12 participants, comparing FlexiCamAR against the baseline, a traditional front-facing AR camera setup, across two common tasks: taking photos and scanning QR codes. Our findings show that FlexiCamAR significantly reduces physical load. We also explore potential scenarios where the additional viewpoint afforded by FlexiCamAR proves valuable, such as capturing low-angle perspectives or navigating confined spaces. Participant feedback further suggests strong potential for additional applications, including selfie taking, video conferencing, and object scanning. Overall, FlexiCamAR presents a novel interaction approach that can serve as a powerful supplement or alternative to the first-person perspective, significantly improving the adaptability of AR glasses for everyday use.
}


\begin{document}



\maketitle
\section{Introduction}
\label{sec:Introduction}
Consumer-grade augmented reality (AR) glasses are emerging as a potential next-generation computing platform, with Mark Zuckerberg claiming that they will gradually replace smartphones~\cite{podcast}. To explore how AR glasses can become everyday computing devices for people, the HCI community has explored a wide range of application scenarios, including social interaction~\cite{janaka2023glassmessaging, rzayev2020effects}, navigation~\cite{lakehal2023spatial}, work~\cite{pavanatto2024working}, entertainment~\cite{simeone2023everyday,zheng2025heythere}, and AI assistance~\cite{cai2025aiget}. A popular design for these glasses is the integration of a front-facing camera that provides a first-person perspective (e.g., Xreal\footnote{https://www.xreal.com/}, Meta\footnote{https://www.meta.com/ai-glasses/}, and Rokid\footnote{https://arstudio.rokid.com/}). This type of front-facing camera supports functions such as environmental detection~\cite{salamone2021wearable,shi2023region} and social sharing~\cite{Nassani2018}. While this approach is intuitive, it fixes the camera to the user's head. As a result, even simple adjustments can require cumbersome head or full-body movements. Smartphones allow users to change viewpoints by moving their hands. However, because the capture point and the viewing point are coupled, users are required to hold the device, look at its screen, and aim the camera at the same time. This coupling can limit flexibility and comfort during everyday camera-based tasks. Therefore, it is necessary to explore alternative approaches that provide the hand-level flexibility of smartphones while preserving the head-worn viewing advantages of AR glasses.

To enhance interaction experiences with AR glasses, a well-explored approach is to augment them with wearable input devices, such as a smart ring~\cite{li2022nailring}, a pen~\cite{zhao2025spatialtouch}, and haptic feedback modules~\cite{gruenefeld2022}. These studies focus on improving the input capabilities of AR systems. For example, they enable new functions such as gesture-based control~\cite{kurz2021smart} and text entry~\cite{li2022nailring}. However, all these studies are based on the traditional first-person perspective. In these studies, the camera is treated as a fixed sensor for the system. Its potential as a flexible tool has not been explored. Similarly, research involving finger-worn devices has explored diverse applications, ranging from gesture interaction~\cite{kurz2021smart, patnaik2025datamancer} and health monitoring~\cite{zhang2019smart} to identity authentication~\cite{cao2024}. When cameras are integrated, they are predominantly treated as sensors for specific recognition tasks, such as context-aware device interaction~\cite{kim2024}, tag sensing~\cite{zhang2020thermalring}, and fabric pattern recognition~\cite{stearns2018applying}. While Stearns et al.~\cite{stearns2017augmented} introduced a controllable viewpoint for low-vision magnification, their design was tailored to assistive reading. Consequently, the use of a flexible, decoupled viewpoint to enhance general-purpose AR interactions remains under-explored.

Our technique, FlexiCamAR, is a new interaction approach for AR glasses that introduces a flexible secondary camera view. It decouples the capture point from the viewing point. Users can move their hands to achieve the desired viewpoint, such as low angles, close-ups, or views inside tight spaces, while maintaining a comfortable posture and viewing the live feed within their AR display (see~\cref{fig:teaser} for some examples). This capability is particularly relevant to AR glasses because it preserves the in-situ, head-worn viewing experience while providing a hand-controlled perspective that is not available with head-fixed cameras. To investigate this approach, we first developed a ring camera prototype as a research tool. We then selected taking photos and scanning QR codes as our evaluation tasks. The selection of these tasks was based on an online survey with 54 responses, which aimed to identify camera-based activities that are frequently performed. While these tasks are intentionally simple and common to provide controlled benchmarks for comparing approaches, a flexible secondary viewpoint can also support richer AR interactions such as spatial inspection, object manipulation assistance, and multi-perspective visualization, as illustrated in~\cref{sec:Unique Use Cases and Scenarios}. Finally, we conducted a formal user study with 12 participants, comparing FlexiCamAR against the baseline, a traditional first-person perspective setup. The study also evaluated how two different display modes influenced user performance and experience.

In short, our main contributions include (1) FlexiCamAR, an approach for AR glasses that uses a flexible secondary camera to enhance users' interaction experiences (see~\cref{sec:approach_and_prototype}); (2) a user study of the physical load compared to the baseline (see~\cref{sec:evaluation}); and (3) an exploration of potential use cases showing how FlexiCamAR enables tasks that are difficult or impossible with a fixed camera (see~\cref{sec:Unique Use Cases and Scenarios}).

\section{Related Work}
\label{sec:Related Work}

\subsection{Wearable Cameras}
\label{sec:wearable_cameras}
Wearable cameras are typically positioned on the torso, head, wrist, or finger. We reviewed the designs and physical constraints of these form factors to inform our approach. Cameras on the torso and head are commonly used for lifelogging~\cite{cartas2020} and healthcare monitoring~\cite{imtiaz2020wearable}. For example, Doherty et al.~\cite{doherty2013using} and Davies et al.~\cite{davies2020feasibility} used hip-mounted cameras combined with accelerometer data to record and analyze everyday physical activities. Similarly, a survey by Allé et al.~\cite{alle2017wearable} reported researchers' successful use of cameras hung around the neck or chest to assist users in documenting and recalling daily events. Additionally, Imtiaz et al.~\cite{imtiaz2020wearable} employed glasses-mounted cameras to monitor users' smoking behaviors. With the development of head-mounted display (HMD) devices, built-in cameras are increasingly used for gesture recognition~\cite{xu2019pointing,shi2023region,ta2024enhancing}. However, the perspectives of these cameras are coupled with the user's body or head movements. This limits their ability to capture viewpoints from confined spaces or diverse angles without uncomfortable body postures.

Wrist-worn cameras offer relatively higher flexibility. Quan et al.~\cite{quan2025} designed a wrist-worn camera that successfully recognized hand gestures under various lighting conditions. However, the wrist offers less dexterity than the fingers. Since fingers are at the end of the hand, they provide the greatest freedom of movement. Consequently, a finger presents a more promising location for flexible camera interaction. Unlike smartphones that occupy the palm and need to be held in specific ways, a ring-based form factor allows users to keep their palms open for other tasks. In terms of specific placement, fingertip-mounted devices like NailRing~\cite{li2022nailring} mount a camera on the fingernail facing the finger pulp, which necessitates a structure that covers the fingertip and tends to obstruct tactile sensation. Stearns et al.~\cite{stearns2017augmented} aligned the camera with the finger's pointing direction to assist with reading tasks. Meanwhile, Kim et al.~\cite{kim2024} placed an upward-facing camera on the dorsal side of a ring for IoT control. Evaluating these designs suggests that the dorsal position is the most suitable for a general-purpose secondary viewpoint. This placement preserves tactile sensitivity better than fingertip-covering designs and maintains grasping capabilities better than palm-occupying devices. Therefore, the dorsal-mounted configuration, similar to the physical form factor in~\cite{kim2024}, emerges as an optimal design choice, providing a natural perspective while preserving hand dexterity.

\subsection{Finger-based Interaction Paradigms}
\label{sec:finger_based_interaction_paradigms}
Existing finger-worn devices primarily utilize sensors as passive input for machine recognition rather than tools for human visual augmentation. Research has extensively explored the use of smart rings for gesture control and context sensing. For example, Zhao et al.~\cite{zhao2021mouse} integrated an embedded IMU sensor into a smart ring to achieve high-precision two-dimensional cursor control on AR displays, while Shen et al.~\cite{shen2024RingGesture} showed that similar sensor integration significantly increases text input efficiency. Beyond IMUs, researchers have explored a range of sensing techniques. Waghmare et al.~\cite{waghmare2023z} employed bio-impedance sensing for robust gesture recognition, and Wang et al.~\cite{wang2024gazering} introduced a pressure-sensitive ring to enhance hand-eye coordination. Additionally, advanced gesture interactions such as bimanual manipulation~\cite{patnaik2025datamancer} and hand-to-face input~\cite{li2023enabling} have been enabled through cross-device sensing frameworks. Similarly, Bardot et al.~\cite{bardot2022one} featured haptic input in a ring to control networked devices.

Some studies have integrated cameras into rings, but the usage paradigm differs fundamentally from our approach. Li et al.~\cite{li2022nailring} developed NailRing, which uses a camera to capture the deformation and color changes of the user's own fingertip for micro-gesture recognition. In this case, the camera is an internal sensor for the system to understand user intent, not an external eye for the user to observe the world. Although concepts like \textit{seeing with the hands} have been explored~\cite{teng2025seeing}, they often rely on sensory substitution rather than direct visual feedback. While Stearns et al.~\cite{stearns2017augmented} utilized a finger-mounted camera to provide visual augmentation, their system was specifically designed to magnify text for users with visual impairments. FlexiCamAR expands this concept to general-purpose AR interaction. We propose a paradigm shift where the wearable camera serves as a user-controlled secondary viewpoint. This active augmentation allows users to intuitively explore the environment, inspect occluded objects, and capture novel perspectives, filling a gap in current finger-based interaction research.

\section{The FlexiCamAR Approach and Prototype}
\label{sec:approach_and_prototype}
In this section, we introduce the core principle of FlexiCamAR. To investigate this approach, we developed a proof-of-concept prototype.

\begin{figure}[tb]
 \centering
 \includegraphics[width=\columnwidth]{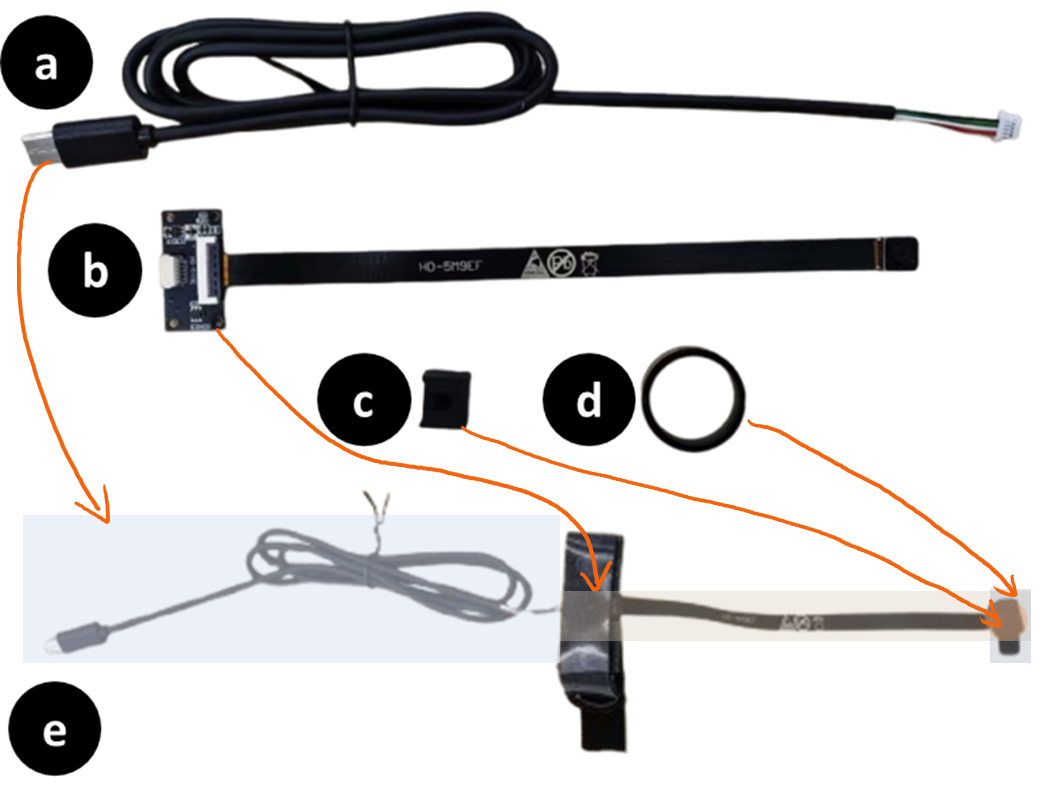}
 \caption{The proof-of-concept prototype and its main components. (a) 1-meter type-C to USB cable link. (b) The camera unit. (c) The 3D-printed enclosure. (d) A size 10 (US standard) ring. (e) The assembled ring camera prototype.}
 \label{fig:ring_camera_hardware}
\end{figure}

\subsection{Prototype Implementation}
\label{sec:prototype_implementation}
The prototype (see~\cref{fig:ring_camera_hardware}) was built by integrating a 1920×1080 pixel, 75° FoV camera unit with a ring component. A 3D-printed enclosure was used to protect the camera unit. The camera module was physically mounted on a finger-worn ring. We used a Rokid Max Pro \footnote{https://arstudio.rokid.com/}, an AR headset supporting six degrees of freedom, as the display and computing platform. For reliable high-frame-rate streaming in the study, the ring camera was wired to the headset via a USB Type-C interface for both power and data, which provided 1920×1080 video at 30 fps with low latency (within 150 ms). In practical deployments, wireless transmission is preferable to maximize freedom of movement, but such a device will only be feasible once miniature components are available. In addition, to enhance stability and comfort, we used auxiliary hook-and-loop fasteners on the user's wrist to fix the camera cable.

\begin{figure}[tb]
 \centering
 \includegraphics[width=\columnwidth]{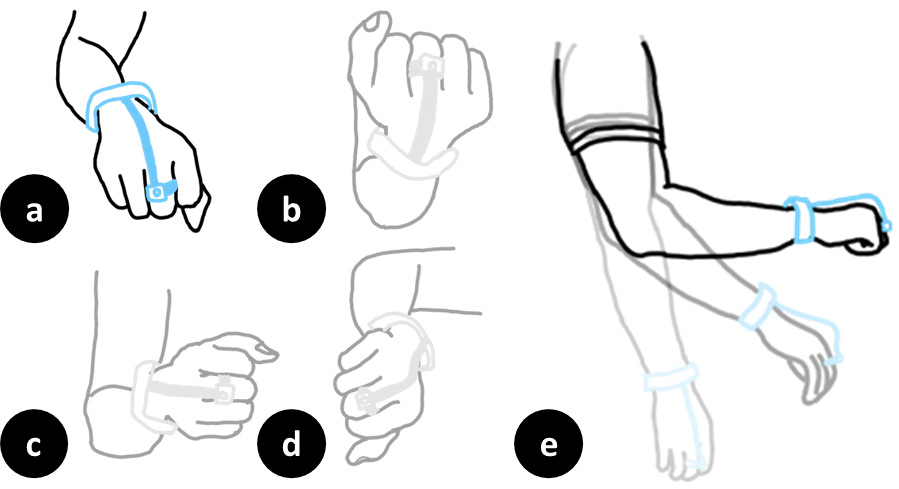}
 \caption{Ergonomic principles for the physical instantiation of FlexiCamAR. (a) The recommended posture, leveraging natural wrist flexion for comfort. (b-d) Postures that were not recommended as they require extra forearm rotation or cause discomfort. (e) The action sequence from a resting state to active use of the approach.}
 \label{fig:pose}
\end{figure}

\begin{figure}[tb]
 \centering
 \includegraphics[width=\columnwidth]{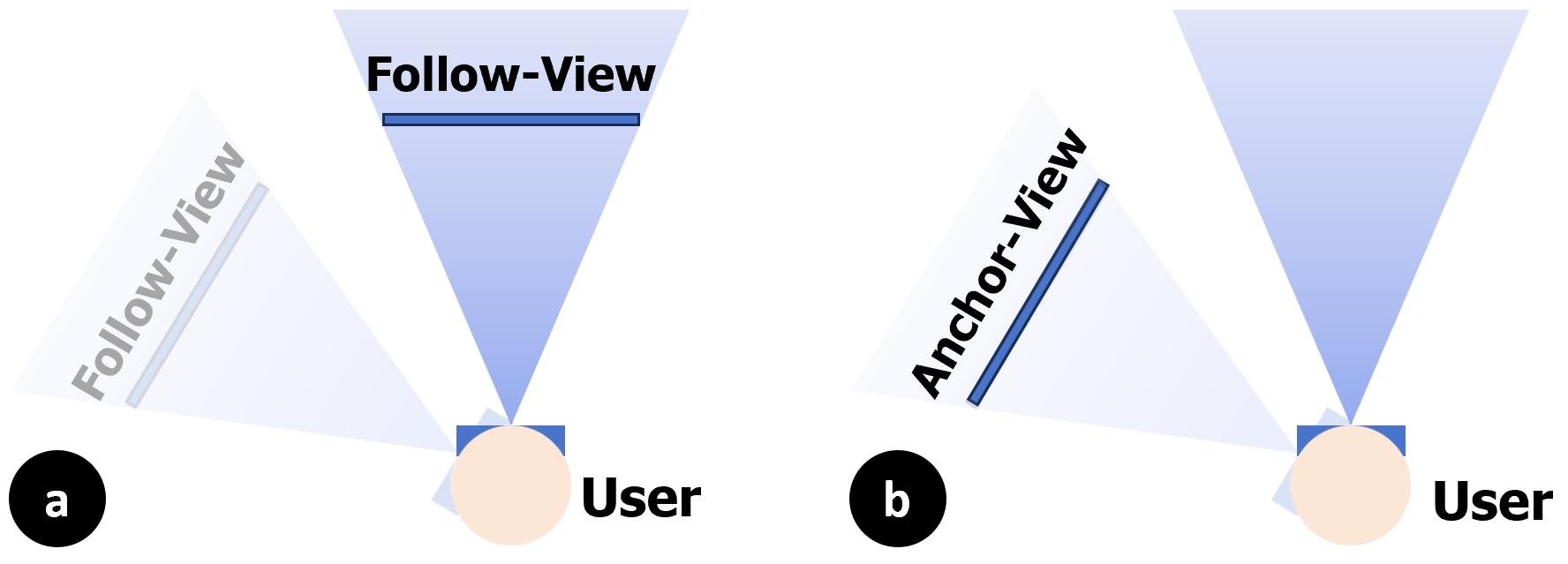}
 \caption{The two display modes explored for presenting the secondary viewpoint of FlexiCamAR. (a) In follow-view, the virtual screen is head-stabilized, following the user's rotation. (b) In anchor-view, the screen is world-stabilized, anchored at a fixed position in mid-air.}
 \label{fig:rotate_view}
\end{figure}

\subsection{Core Principles of FlexiCamAR}

\begin{figure*}[tb]
  \centering
 \includegraphics[width=\textwidth]{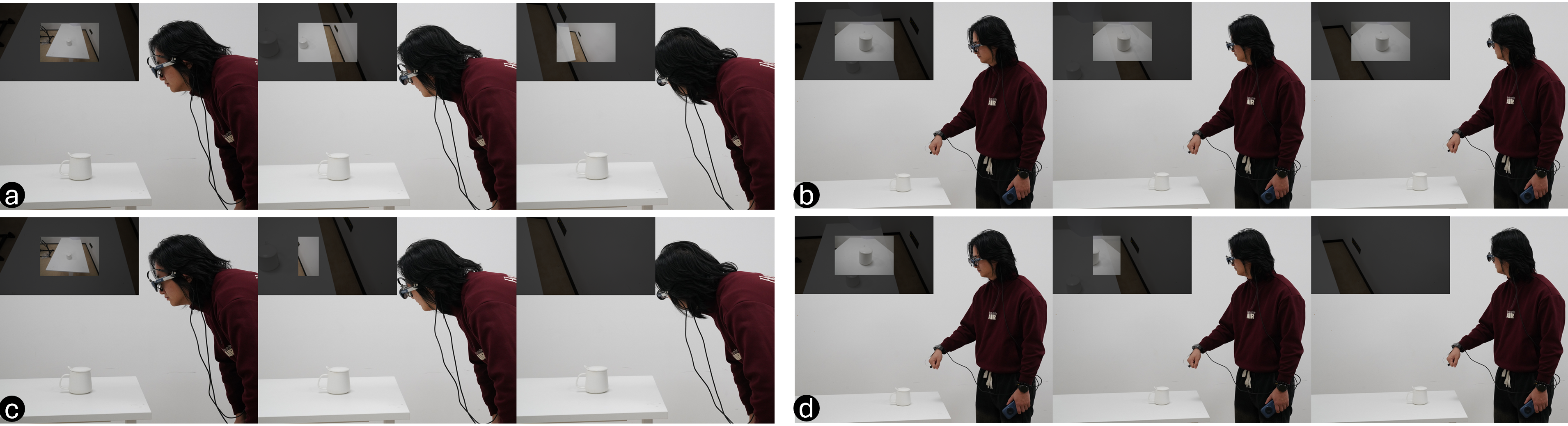}
 \caption{Visual illustrations of the four experimental conditions. (a) Baseline with Follow-View. (b) FlexiCamAR with Follow-View. (c) Baseline with Anchor-View. (d) FlexiCamAR with Anchor-View.}
 \label{fig:sim_view}
\end{figure*}

\label{sec:core_principles}
A core principle of FlexiCamAR is to enhance interaction flexibility while maintaining user comfort. This principle is achieved by decoupling the camera's viewpoint from the user's head, leveraging the natural dexterity of the hand and its wide range and flexibility of motions. To find the most comfortable design, we conducted formative interviews with six experienced AR glasses users (4 males and 2 females; aged between 25--32, $mean=27.5, SD=2.63$). We recruited experienced users to ensure that feedback was grounded in a realistic understanding of AR limitations, such as field of view (FoV) and device weight, rather than novelty effects from first-time use. All participants had at least one year of experience with AR/VR headsets. During the interview, participants were asked to perform mock camera tasks (e.g., shooting at arbitrary angles and aiming for QR code scanning)  using a camera module and verbalize their preferred mounting positions and display modes. Participants started from a relaxed state (sitting or standing) and transitioned to a "working posture." They were not told to use specific hand gestures. Most participants naturally adopted a clenched or slightly clenched fist, while two participants employed a relaxed palm posture (similar to the intermediate state in \cref{fig:pose}.e) for simulated low-angle shots. The sessions were audio-recorded and transcribed for analysis. Based on the feedback of the participants, the ideal placement of this approach is to place the camera on the middle finger with the palm facing downward when the secondary viewpoint is horizontally upright (as shown in~\cref{fig:pose}a). The middle finger is selected due to the stabilization, especially when it is clamped together with the index and ring fingers. Furthermore, orienting the palm downward leverages the natural flexibility of wrist flexion and extension, offering enhanced comfort compared to lateral wrist movements~\cite{moser2020importance}. Since people usually rest with their palms facing downward or toward their bodies, this usage posture enables users to capture images by simply lifting the arm without extra rotation (see~\cref{fig:pose}e). Our study primarily focused on tasks aligned with the user's direct line of sight, such as photographing and scanning QR codes, making the camera placement on the back of the hand more suitable. For tasks that require images in the opposite direction (e.g., selfies), placing the camera on the palm side might be more comfortable. Since our study focused on camera interaction, we used the standard Rokid controller for all other interactions. However, this aspect could be replaced with alternative devices or mechanisms depending on the specific features of the AR HMD.

A second core aspect of FlexiCamAR is the presentation of the secondary viewpoint. To ground our design, we referred to established theoretical frameworks for AR information spaces~\cite{billinghurst1998evaluation, ens2014ethereal}, which distinguish between head-referenced (egocentric) and body/world-referenced (allocentric) frames. FlexiCamAR applies these concepts by decoupling the camera from the head. This allows users to benefit from the viewing comfort of a head-referenced display while utilizing the flexibility of hand-referenced camera manipulation.

In selecting the display modes, we considered various configurations explored in prior work, such as body-referenced menus~\cite{lediaeva2020evaluation}, finger-augmented displays~\cite{stearns2017augmented}, and multi-display arrangements~\cite{ha2006my, das2010evaluation}. However, we excluded body- or finger-referenced modes for this study to ensure fairer comparative evaluation. For the baseline (front-facing camera), viewing a display attached to the hand would force the user to look away from the scene, causing the camera to lose its target. We also excluded object-referenced modes~\cite{cheng2021semanticadapt} to ensure the system works in any environment, regardless of the surroundings. Finally, we did not use adaptive placement~\cite{lu2024adaptive}. Although adaptive algorithms are promising, they can introduce uncertainty if they misinterpret the user's intent. To clearly evaluate the interaction technique itself, we chose consistent and predictable display modes.

Consequently, we selected two fundamental modes that are standard in commercial AR glasses (e.g., XREAL One Pro\footnote{\url{https://www.xreal.com/one-pro}}) and were endorsed by participants in our formative interviews:
\begin{description}
    \item[Follow-View] (Head-referenced) A virtual screen that follows the user's head rotation, maintaining constant visibility directly ahead (see~\cref{fig:rotate_view}a).
    \item[Anchor-View] (World-referenced) A virtual screen anchored at a fixed location in mid-air, reducing distraction while providing the option to be repositioned (see~\cref{fig:rotate_view}b).
\end{description}

Therefore, to understand how these display modes influence user experience across different camera configurations, we incorporated them into our evaluation. This resulted in four experimental conditions combining the two interaction approaches (Baseline and FlexiCamAR) with the two display modes, as illustrated in~\cref{fig:sim_view}.

\section{Evaluation}
\label{sec:evaluation}
We conducted a user study to evaluate FlexiCamAR. To guide our investigation, we formulated the following two research questions:
\begin{itemize}
    \item \textbf{RQ1:} How does the decoupled viewpoint (FlexiCamAR) compare to the coupled viewpoint (Baseline) in terms of efficiency, physical load, and user experience?
    \item \textbf{RQ2:} How does the display reference frame (Follow-View vs. Anchor-View) impact the usability and performance of these interaction techniques?
\end{itemize}

\begin{figure}[tb]
 \centering
 \includegraphics[width=0.8\columnwidth]{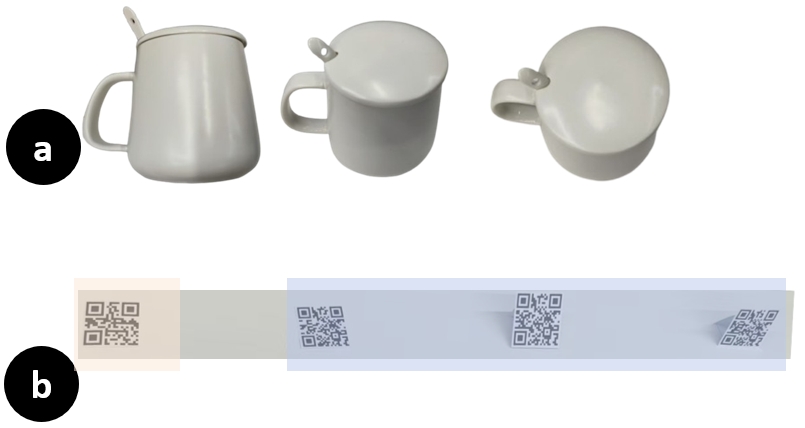}
 \caption{The experiment tasks. (a) The photo-taking task: Participants take photos of a cup from three angles. (b) The QR code-scanning task: Participants scan the QR codes from left to right. The QR code in the red block is the start marker, and the QR codes in the blue block are the task codes.}
 \label{fig:environment}
\end{figure}

\subsection{Tasks}
\label{sect:tasks}
We aimed to evaluate FlexiCamAR in everyday tasks. To choose suitable experimental tasks, we first distributed a questionnaire to survey users' most frequently performed activities using their smartphone cameras, which are more commonly used in their daily lives. The results of 54 respondents showed that the two most frequently performed activities using smartphone cameras were taking photos and scanning QR codes. Thus, we used these two as our experimental tasks in this study.

The photo-taking and QR code-scanning tasks were conducted in the same room to keep environmental factors consistent. The room had a table with a height of 75 cm, around which the participants could move freely within a 1 meter radius.

In the photo-taking task, participants were required to take photos of a cup placed at the center of the table from three angles: level, oblique, and top-down (shown in~\cref{fig:environment}a). Participants were not restricted in their shooting posture. When the shutter button was pressed, the AR glasses played an audio effect, and the photo remained on the AR glasses for one second for the participants to review.

In the QR code-scanning task, four QR codes were arranged side-by-side on the table with a spacing of 20 cm between each (shown in~\cref{fig:environment}b). The participants began the task by scanning the QR code on the left, which served as a start marker. The participants were then asked to scan the flat, tilted (at 45°), and vertical (tilted at 90°) QR codes sequentially from left to right. We selected the table-top setting because it is common in daily life (e.g., reading documents, scanning payment codes on counters). When a QR code entered the camera's view, the system automatically attempted to recognize it, and an audio effect was played only after successful recognition, prompting participants to proceed to the next QR code.

\begin{figure}[tb]
 \centering
 \includegraphics[width=0.5\columnwidth]{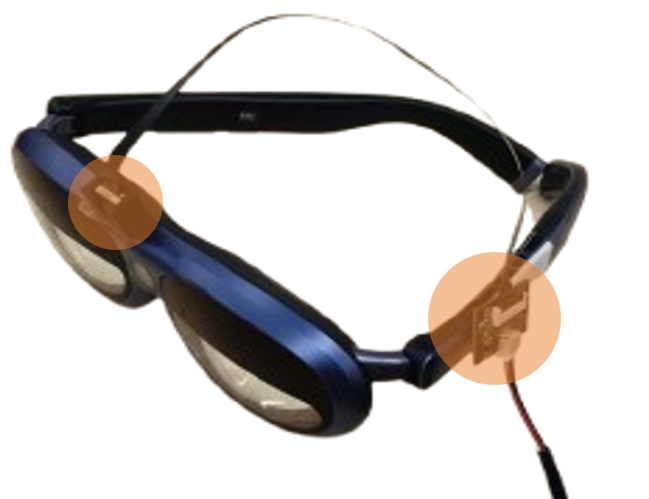}
 \caption{The experimental setup for the baseline. A camera unit was attached to the front of the AR headset to create the fixed, first-person perspective for our comparative study.}
 \label{fig:ar_front-facing_camera}
\end{figure}

\subsection{Apparatus}
\label{sec:apparatus}
FlexiCamAR was realized together with the Rokid AR headset. The baseline replicated the default input of AR glasses (i.e., a front-facing camera) on the same headset (see~\cref{fig:ar_front-facing_camera}). This design keeps the comparison within AR techniques and evaluates whether a secondary viewpoint improves everyday camera interactions on AR glasses, rather than comparing AR with smartphones. Since the original front camera of the Rokid AR headset is black and white, this extra camera was necessary to capture colored images. The extra camera connected to the AR glasses was identical to the one on the finger-worn camera prototype for FlexiCamAR. During the experiment, the data captured by users was sent over the network to a remote computer for storage.

The camera feed was displayed on the AR glasses as a virtual screen (width: 0.32 m, height: 0.18 m) positioned at a virtual distance of 1 m. This screen was rendered as an overlay, ensuring it remained visible and was not occluded by real-world objects. In the \textit{anchor-view} mode, the screen was world-stabilized at a 1-meter distance but could be instantly recalibrated to the user's front view. For the QR code-scanning task, the codes were generated using the Python \texttt{qrcode} library (pixel size: 10$\times$10) and recognized using the OpenCV library (Python \texttt{cv2}).

\subsection{Participants}
\label{sec:participants}
We recruited 12 participants (6 males and 6 females) from a local university community. Their age ranged from 22 to 32 years ($mean = 26.9, SD = 3.04$) and their heights ranged from 160 to 180 cm ($mean = 170.5, SD = 6.78$). All participants were right-handed. Four participants had prior experience with AR glasses, while the remaining eight had no prior experience. Each participant received approximately 8 USD for their participation. All participants provided written informed consent. They were informed in detail about the types of data recorded, the recording methods, and the duration of data storage. This study was reviewed and approved by the Human Subjects Panel of the Hong Kong University of Science and Technology (Guangzhou) under ethics application number HKUST(GZ)-HSP-2025-0114.

\subsection{Study Design and Procedure}
\label{sec:study_design}
To evaluate the proposed FlexiCamAR approach, we employed a within-subjects study design to compare its performance and user experience against the baseline. The study had two independent variables: the \textsc{Interaction\_approach} and the \textsc{Display} mode. The \textsc{Interaction\_approach} had two levels: (a) \textbf{FlexiCamAR}, realized via the finger-worn camera prototype, and (b) the \textbf{baseline}, realized via the front-facing camera setup. The \textsc{Display} mode of the camera view also had two levels: \textit{follow-view} and \textit{anchor-view}. This 2$\times$2 design resulted in four experimental conditions: (1) Baseline + follow-view; (2) Baseline + anchor-view; (3) FlexiCamAR + follow-view; and (4) FlexiCamAR + anchor-view.

The experiment began by explaining the experimental procedure to participants and obtaining their informed consent. The evaluation was split into two sessions by tasks. In each session, participants were asked to complete the given task under all four conditions, the order of which was counterbalanced using a balanced Latin square approach. In the photo-taking task, participants used each technique to take photos from three angles five times; while in the QR code-scanning task, participants used each technique to scan the four QR codes five times. Participants were given a 5-minute break between the two sessions. After completing each session, participants were asked to evaluate the techniques through questionnaires. We also conducted semi-structured interviews focused on perceived physical effort, comfort of the ring form factor, and preference between display modes. These sessions were audio-recorded and transcribed for further analysis. After participants completed all tasks, they were asked to provide additional feedback on the interaction approaches. The entire experiment lasted approximately 40 minutes per participant.

\subsection{Metrics}
\label{sec:metrics}
We collected both objective and subjective data to answer the research questions. For \textbf{RQ1} (efficiency and physical load), we automatically recorded task completion times. Completion time was defined as the average time to take a photograph in the photo-taking task and the average time to achieve a successful QR code recognition in the scanning task. We assessed workload using the NASA Task Load Index (NASA-TLX)~\cite{hart2006nasa}. The NASA-TLX includes six subscales: Mental Demand, Physical Demand, Temporal Demand, Performance, Effort, and Frustration.

For \textbf{RQ1} (user experience) and \textbf{RQ2} (impact of display modes), we measured perceived usability with the System Usability Scale (SUS)~\cite{brooke1996sus} and user experience with the User Experience Questionnaire-Short (UEQ-S)~\cite{schrepp2017design} across the interaction approaches and display modes. Semi-structured interviews provided qualitative context for interpreting both RQs.

\section{Results and Discussion}
\label{sec:Results and Discussion}
The collected data was analyzed using the two-way repeated measures analysis of variance (RM-ANOVA). Furthermore, the subjective data were transformed using Aligned Rank Transform~\cite{wobbrock2011aligned} before conducting RM-ANOVA.

\begin{figure}[tb]
 \centering
 \includegraphics[width=\columnwidth]{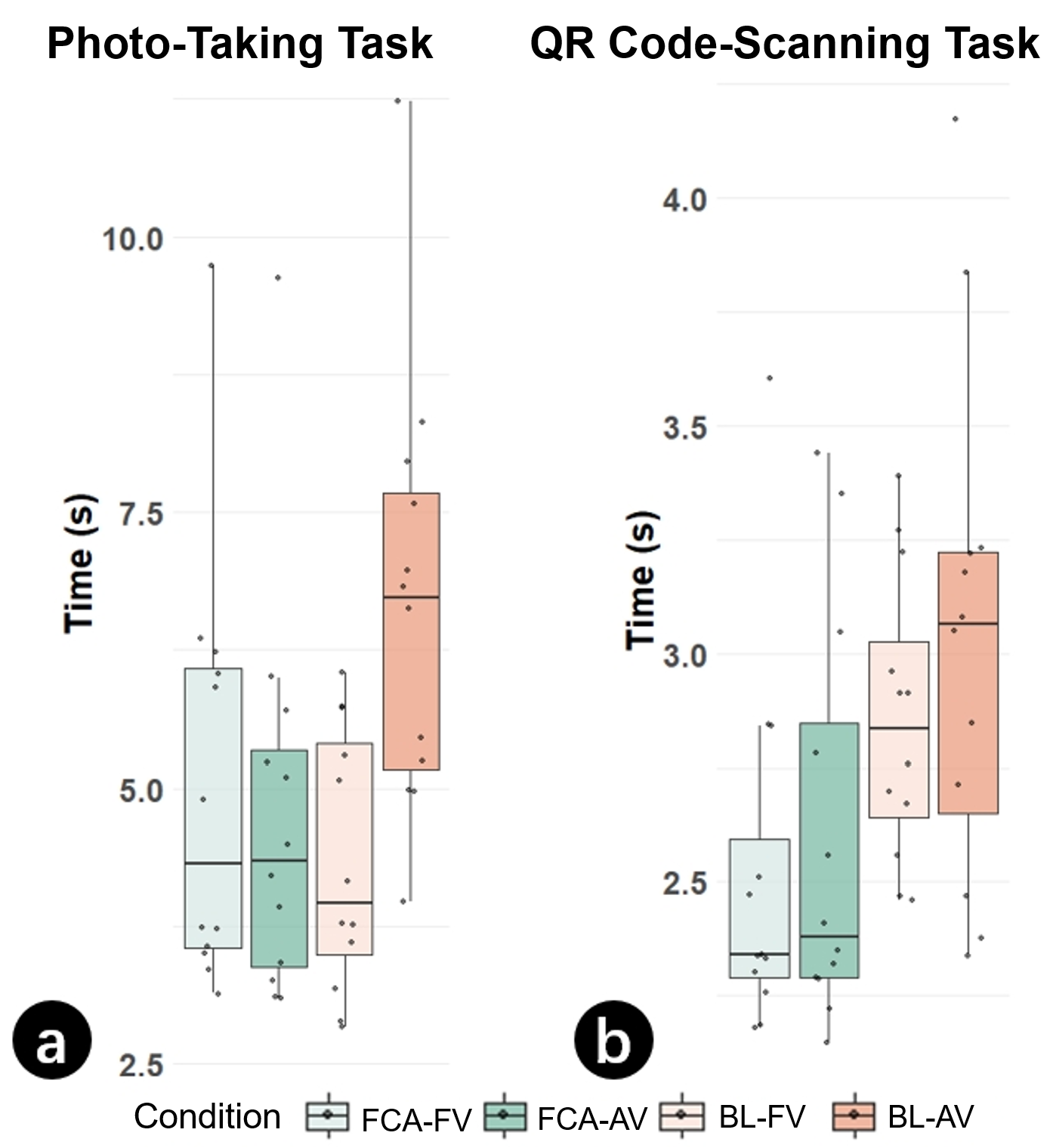}
 \caption{Average task completion time across the two interaction approaches and two display modes, for (a) the Photo-taking task and (b) the QR code-scanning task. (Legend: FCA: FlexiCamAR; BL: baseline; FV: Follow-View; AV: Anchor-View).}
 \label{fig:time_analysis}
\end{figure}

\subsection{Average Completion Time}
\label{sec:results_time}
\Cref{fig:time_analysis} shows the average task completion time under the four \textsc{Interaction\_approach} $\times$ \textsc{Display} conditions for the two tasks. The collected data was analyzed using two-way RM-ANOVA.

For the photo-taking task, RM-ANOVA tests revealed that the main effect of \textsc{Interaction\_approach} was not statistically significant ($F(1, 33)=2.96, p=.095$). However, the main effect of the \textsc{Display} was significant ($F(1, 33)=8.33, p=.007$). Additionally, there was a significant interaction effect between \textsc{Interaction\_approach} and \textsc{Display} ($F(1, 33) = 12.94, p = .001$). Post-hoc tests indicated that when using the baseline, the follow-view ($mean = 4.34 s, SD = 1.17$) was significantly faster than the anchor-view ($mean = 6.68 s, SD = 1.97; p < .001$). Conversely, we found no significant difference between the two display modes on average completion time when using FlexiCamAR ($p=.682$). We also found that in the anchor-view mode, FlexiCamAR enabled significantly faster task completion ($mean = 4.76 s$) than the baseline ($mean = 6.68 s; p < .001$).

For the QR code-scanning task, RM-ANOVA tests revealed a significant main effect of \textsc{Interaction\_approach} ($F(1, 33)=10.86, p=.002$), indicating that FlexiCamAR ($mean = 2.56 s$) enabled significantly faster task completion than the baseline ($mean = 2.95 s; p=.020$). However, the main effect of the \textsc{Display} ($F(1, 33)=1.30, p=.262$) and the interaction effect ($F(1, 33)=0.19, p=.669$) were not statistically significant.

In summary, for the photo-taking task, our results suggest a key difference in the nature of the two approaches: the performance of the baseline is highly dependent on the choice of display mode, whereas FlexiCamAR demonstrates greater robustness across different display conditions. The benefit of follow-view over anchor-view was only found with the baseline. For the QR code-scanning task, which requires rapid and frequent viewpoint adjustments, FlexiCamAR consistently demonstrated a significant performance advantage over the baseline, irrespective of the display mode used. And the display mode itself did not significantly impact task completion time.

\begin{figure}[tb]
 \centering
 \includegraphics[width=\columnwidth]{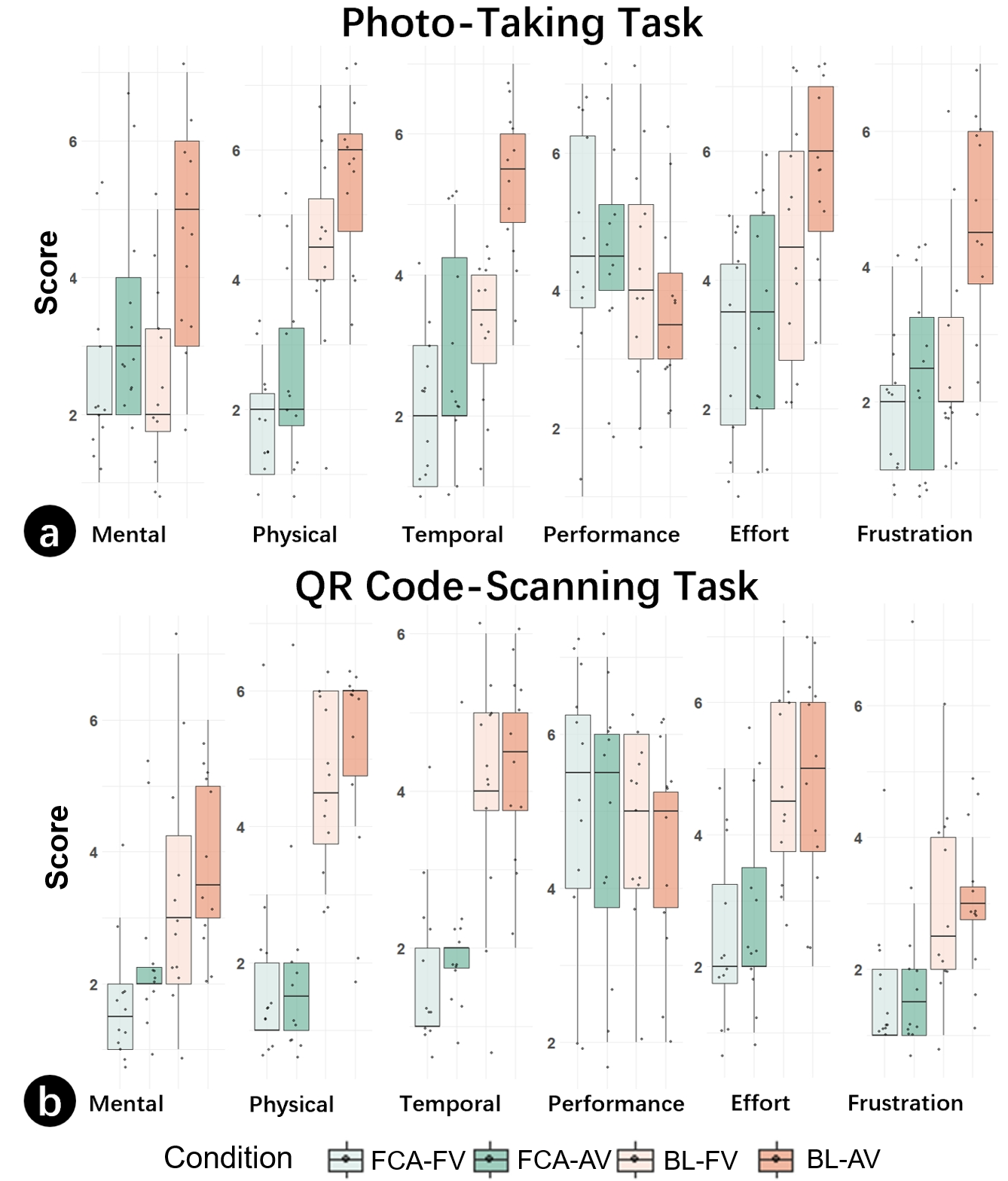}
 \caption{NASA-TLX scores of the four experimental conditions in (a) the Photo-taking task and (b) the QR code-scanning task. FCA: FlexiCamAR; BL: baseline; FV: Follow-View; AV: Anchor-View.}
 \label{fig:nasa_analysis}
\end{figure}

\begin{table*}[htbp]
    \centering
    \begin{tabular}{lllll}
        \toprule
        Task            & Dimensions  & \textsc{Interaction\_approach}      & \textsc{Display}                      & \textsc{Interaction\_approach} × \textsc{Display} \\
        \midrule
        Photo-Taking    & Mental      & $F(1,33)=3.47, p=.072$            & \cellcolor{deeporange}$F(1,33)=23.25, p<.001$   & \cellcolor{lightorange}$F(1,33)=4.17, p=.049$  \\
                        & Physical    & \cellcolor{deeporange}$F(1,33)=46.00, p<.001$  & \cellcolor{lightorange}$F(1,33)=4.40, p=.044$    & $F(1,33)=0.28, p=.601$  \\
                        & Temporal    & \cellcolor{deeporange}$F(1,33)=27.84, p<.001$  & \cellcolor{deeporange}$F(1,33)=14.26, p<.001$    & $F(1,33)=3.68, p=.064$  \\
                        & Performance & \cellcolor{lightorange}$F(1,33)=5.17, p=.030$  & $F(1,33)=1.03, p=.318$                  & $F(1,33)=0.55, p=.462$  \\
                        & Effort      & \cellcolor{deeporange}$F(1,33)=19.52, p<.001$  & $F(1,33)=3.69, p=.064$                  & $F(1,33)=1.01, p=.321$  \\
                        & Frustration & \cellcolor{deeporange}$F(1,33)=14.79, p<.001$  &  \cellcolor{medorange}$F(1,33)=11.68, p=.002$   & \cellcolor{lightorange}$F(1,33)=5.36, p=.027$  \\
        \midrule
        QR Code-Scanning & Mental      & \cellcolor{deeporange}$F(1,33)=36.91, p<.001$  & \cellcolor{lightorange}$F(1,33)=4.94, p=.033$    & $F(1,33)=0.01, p=.911$    \\
                        & Physical    & \cellcolor{deeporange}$F(1,33)=37.62, p<.001$  & $F(1,33)=1.90, p=.177$                  & $F(1,33)=0.24, p=.627$    \\
                        & Temporal    & \cellcolor{deeporange}$F(1,33)=59.45, p<.001$  & $F(1,33)=1.06, p=.311$                  & $F(1,33)=0.04, p=.851$    \\
                        & Performance & $F(1,33)=4.03, p=.053$                  & $F(1,33)=0.84, p=.367$                  & $F(1,33)=0.00, p=1.000$   \\
                        & Effort      & \cellcolor{deeporange}$F(1,33)=24.97, p<.001$  & $F(1,33)=0.17, p=.686$                  & $F(1,33)=0.01, p=.919$    \\
                        & Frustration & \cellcolor{deeporange}$F(1,33)=31.97, p<.001$  & $F(1,33)=0.11, p=.744$                  & $F(1,33)=0.31, p=.579$    \\
        \bottomrule
    \end{tabular}
    \caption{RM-ANOVA test results for NASA-TLX scores.}
    \label{tab:nasa}
\end{table*}

\subsection{NASA-TLX}
\label{sec:results_tlx}
\Cref{fig:nasa_analysis} shows the NASA-TLX scores across the four \textsc{Interaction\_approach} $\times$ \textsc{Display} conditions for the two tasks. Significant main effects of \textsc{Interaction\_approach} and \textsc{Display} were observed across several NASA-TLX dimensions, as summarized in~\Cref{tab:nasa}. Interaction effects were only found in mental ($F(1,33)=4.17, p=.049$) and frustration ($F(1,33)=5.36, p=.027$) dimensions in the photo-taking task. This pattern suggests that the baseline's mental demand and frustration are highly sensitive to the display mode, whereas FlexiCamAR maintains a stable workload regardless of the display condition.

In the photo-taking task, participants generally perceived a lower workload when using FlexiCamAR compared to the baseline. Compared to the baseline, FlexiCamAR induced significantly lower physical workload ($p<.001$), temporal workload ($p=.001$), effort ($p=.006$), and frustration ($p=.005$) in the anchor-view mode, and lower physical workload ($p=.001$) and temporal workload ($p=.023$) in the follow-view mode. From another perspective, when using the baseline, participants rated a lower perceived workload with the follow-view than the anchor-view across the mental ($p<.001$), physical ($p=.010$), temporal ($p<.001$), and frustration ($p<.001$) dimensions. This effect was less pronounced with FlexiCamAR, where it was only significant for the mental ($p=.017$) and frustration ($p=.024$) dimensions.

In the QR code-scanning task, FlexiCamAR resulted in significantly lower workload ratings than the baseline in all dimensions except for performance, with both the follow-view (all $p<=0.002$) and the anchor-view (all $p<=0.018$). When comparing the two \textsc{Display} modes, we only found that the follow-view led to a significantly lower mental workload than the anchor-view ($p<0.05$ for both approaches).

In summary, the NASA-TLX results provide strong evidence for the primary ergonomic advantage of FlexiCamAR: its ability to significantly reduce physical load, temporal pressure, and overall effort for the user. This advantage was particularly pronounced in the QR code-scanning task, suggesting that the benefits of FlexiCamAR are magnified in tasks that require rapid, frequent, and fine-grained viewpoint adjustments.

\begin{figure}[tb]
 \centering
 \includegraphics[width=0.5\columnwidth]{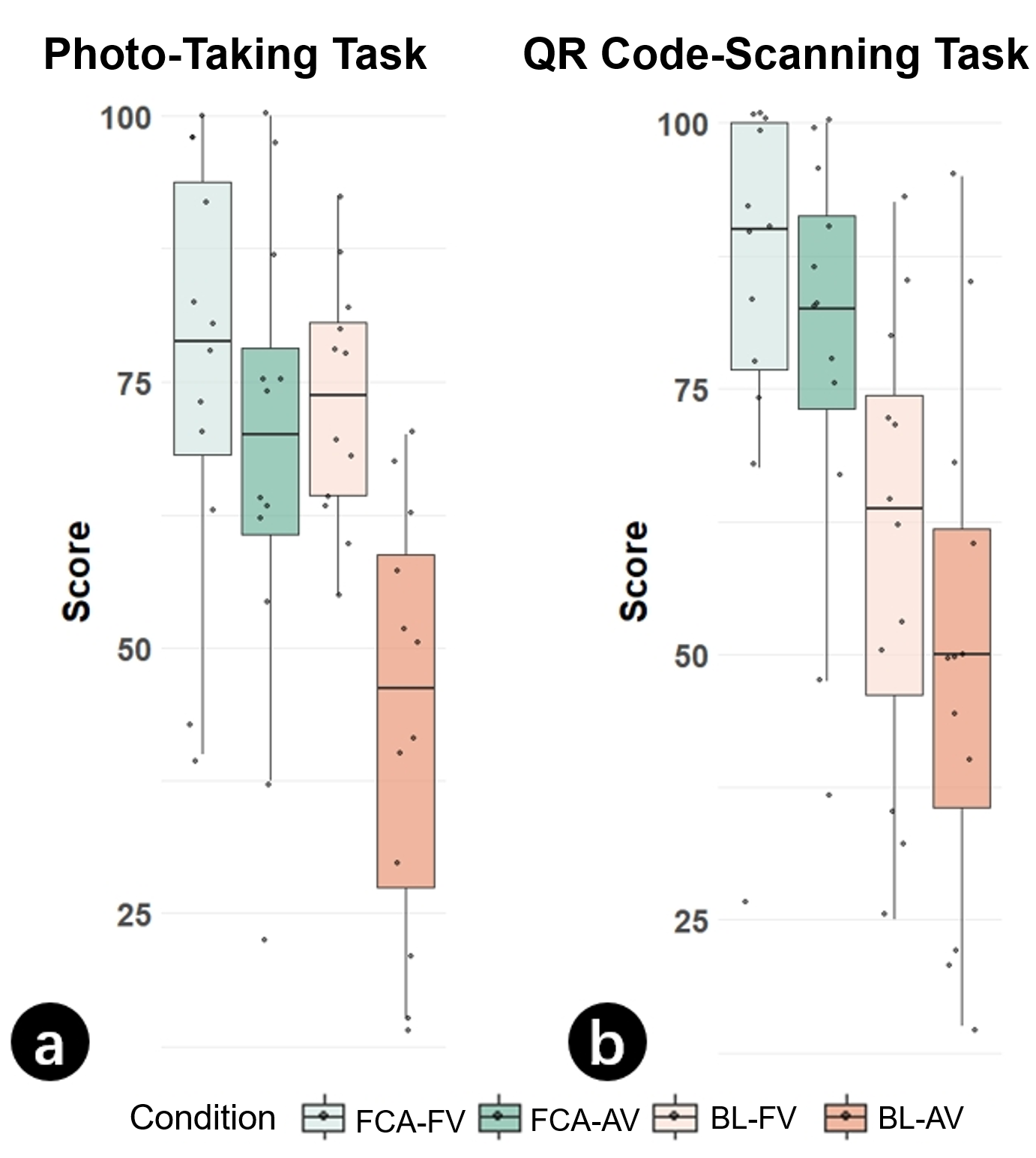}
 \caption{The SUS scores for the four experimental conditions in (a) the Photo-taking task and (b) the QR code-scanning task. FCA: FlexiCamAR; BL: baseline; FV: Follow-View; AV: Anchor-View.}
 \label{fig:SUS_analysis}
\end{figure}

\subsection{SUS}
\label{sec:results_sus}
\Cref{fig:SUS_analysis} presents the SUS scores across the four \textsc{Interaction\_approach} $\times$ \textsc{Display} conditions for the two tasks. In the photo-taking task, significant effects were found for both the \textsc{Interaction\_approach} ($F(1,33)=8.17, p=.007$) and \textsc{Display} ($F(1,33)=11.64, p=.002$) variables, whereas the interaction did not reach significance ($F(1,33)=2.92, p=.097$). In the QR code-scanning task, only the \textsc{Interaction\_approach} variable had a significant effect on SUS scores ($F(1,33)=20.68, p<.001$), while the \textsc{Display} variable ($F(1,33)=2.80, p=.104$) and the interaction effect ($F(1,33)=0.41, p=.526$) showed no significant differences. 

In the photo-taking task, post-hoc analyses show that the perceived usability of FlexiCamAR ($mean = 67.9$) was rated significantly higher than that of the baseline ($mean = 43.5, p = .014$) when the \textsc{Display} mode was anchor-view. However, no significant difference was observed between the two approaches under the follow-view mode ($p = .422$). For the baseline, perceived usability was significantly lower in the anchor-view mode ($mean = 43.5$) compared to the follow-view mode ($mean = 73.1, p < .001$). In contrast, this difference was not significant for FlexiCamAR ($p = .182$). In the QR code-scanning task, post-hoc analyses demonstrated that FlexiCamAR received significantly higher usability ratings in both follow-view ($mean = 83.5$) and anchor-view ($mean = 78.5$) modes than the baseline in the corresponding follow-view ($mean = 60.4, p < .010$) and anchor-view ($mean = 50.0, p < .010$) modes, respectively.

In summary, the SUS scores highlight a key characteristic of the two approaches' usability: FlexiCamAR offers a robust and consistent user experience in different display modes. In contrast, the usability of the baseline appears to be fragile and highly dependent on the display mode, degrading significantly when the view is not constantly centered. FlexiCamAR also showed superior usability in the QR code-scanning task, confirming its advantage for tasks requiring interactional flexibility.

\begin{figure}[tb]
 \centering
 \includegraphics[width=\columnwidth]{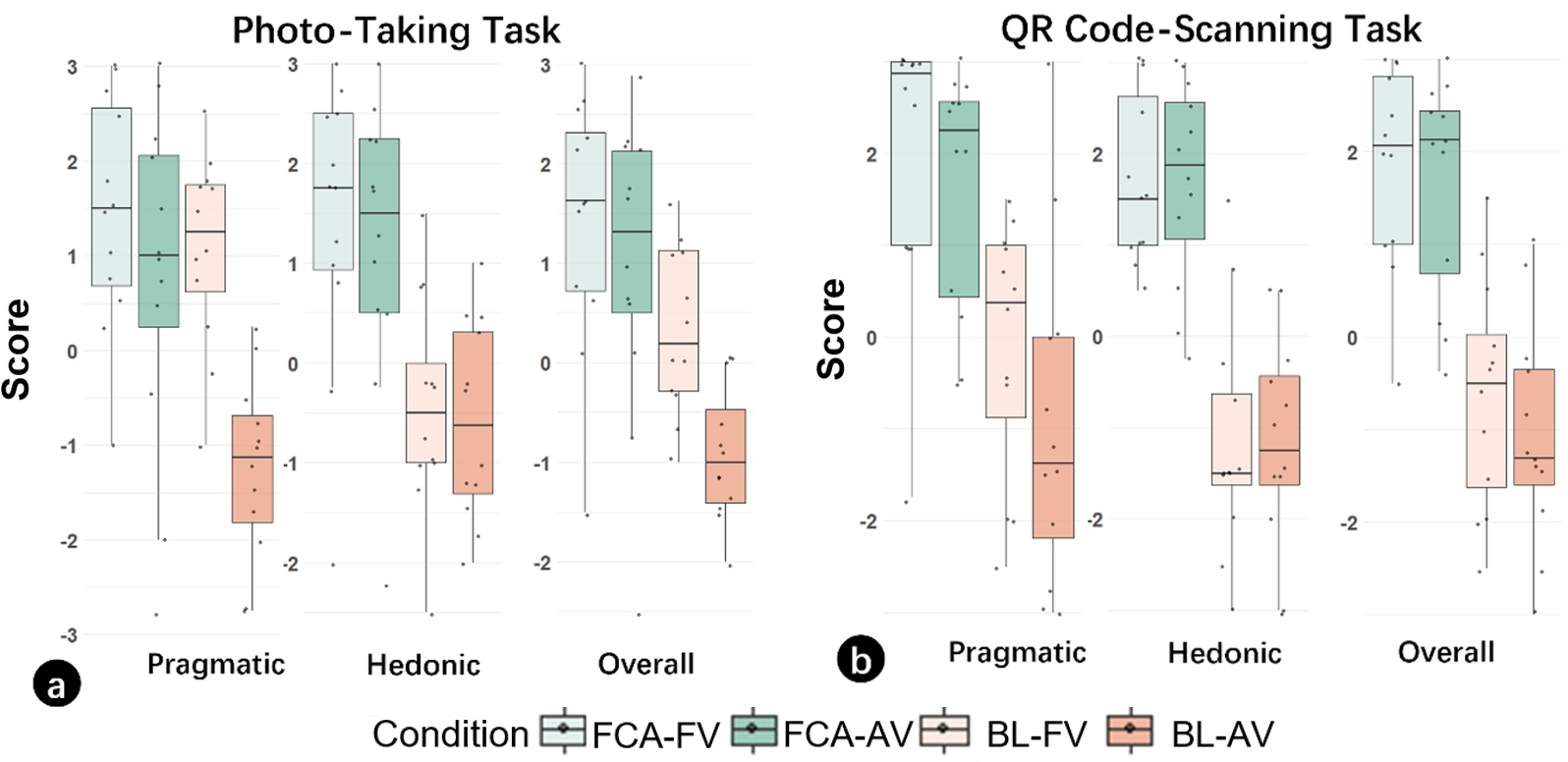}
 \caption{The UEQ-S scores for the four experimental conditions in (a) the Photo-taking task and (b) the QR code-scanning task. FCA: FlexiCamAR; BL: baseline; FV: Follow-View; AV: Anchor-View.}
 \label{fig:UEQ_analysis}
\end{figure}

\subsection{UEQ-S}
\label{sec:results_ueq}
\Cref{fig:UEQ_analysis} presents the UEQ-S scores for various dimensions across the four \textsc{Interaction\_approach} $\times$ \textsc{Display} conditions. For the photo-taking task, \textsc{Interaction\_approach} significantly influenced pragmatic quality ($F(1,33)=12.34, p=.0013$), hedonic quality ($F(1,33)=35.50, p<.001$), and overall score ($F(1,33)=26.77, p<.001$); \textsc{Display} affected pragmatic quality ($F(1,33)=17.76, p<.001$) and overall score ($F(1,33)=8.60, p=.0061$); the interaction was only significant for pragmatic quality ($F(1,33)=6.47, p=.0159$). For the QR code-scanning task, \textsc{Interaction\_approach} showed significant effects across all dimensions (all $p<.001$), while \textsc{Display} significantly affected pragmatic quality ($F(1,33)=5.94, p=.020$) only, and the interaction did not show any significant effects.

In the photo-taking task, post-hoc analyses revealed that FlexiCamAR consistently received higher UEQ ratings than the baseline on the hedonic and overall qualities with both \textsc{Display} modes (all $p<=.003$). Regarding pragmatic quality, FlexiCamAR was rated significantly higher than the baseline in the anchor-view only ($p=.006$). On the other hand, when using the baseline, the follow-view achieved a better user experience than the anchor-view on pragmatic and overall qualities (both $p<.001$).

In the QR code-scanning task, FlexiCamAR consistently received higher ratings than the baseline on all sub-dimensions with both follow-view and anchor-view (all $p<.001$). No other significant differences were found in the post-hoc tests.

In summary, the UEQ-S results indicate that FlexiCamAR offers a comprehensively superior user experience compared to the baseline. This superiority is multifaceted. In the photo-taking task, its primary advantage was in the hedonic dimension, suggesting that the flexibility and novelty of the approach are perceived as more enjoyable and stimulating. In the QR code-scanning task, FlexiCamAR was again better than the front-facing camera across all quality dimensions, regardless of \textsc{Display} modes. The effects of \textsc{Interaction\_approach} were highly significant, while the \textsc{Display} did not significantly impact hedonic or overall quality. These results suggest that FlexiCamAR offers a more consistent and stable user experience across interaction settings.

\subsection{Interview Comments}
\label{sec:results_interview}
Based on the analysis of the semi-structured interviews, FlexiCamAR was favored by most participants in terms of overall preference. For the photo-taking task, 8 out of 12 participants ranked it (either the follow-view or the anchor view) as their top choice. This preference was even stronger in the QR code-scanning task. FlexiCamAR was ranked first by 11 out of 12 participants. Participants attributed their preference to the core ergonomic benefits of FlexiCamAR: its high degree of operational flexibility and a significantly reduced physical burden. Specifically, several participants noted that only slight adjustments to the hand and wrist were required to control the viewing angle, significantly decreasing fatigue and improving overall comfort (P1, P4, P5, P6, P8, P10, P12). However, three participants reported discomfort because they were not used to directly adjusting the viewing angle by hand movements (P3, P7, P11). This suggests that it takes time for people to adapt to a novel interaction approach that decouples hand movement from head movement.

Regarding the display modes within FlexiCamAR, the follow-view was more popular than the anchor-view. As P1 explained, the follow-view mode "\textit{always kept the target centered within the field of view, thus minimizing the likelihood of losing sight of it}." In contrast, the anchor-view mode was considered cumbersome due to the additional calibration required (P1, P3, P8, P10). Overall, participants believed that FlexiCamAR demonstrated unique advantages, particularly in scenarios that are ergonomically challenging for the baseline, such as capturing low-angle shots or inspecting narrow spaces (P1, P4, P5, P9, P10), which we explore further in~\cref{sec:Unique Use Cases and Scenarios}.

Regarding the baseline, although some participants (P3, P7, P11) acknowledged its advantage of providing a stable viewpoint in the follow-view mode, it was generally less favored. This was primarily due to its drawbacks, including limited flexibility in adjusting the viewing angle and higher physical exertion required during use (P1, P2, P4, P5, P6, P8, P10, P12).

In conclusion, participants consistently identified FlexiCamAR, particularly in follow-view mode, as the best choice for scenarios involving frequent minor adjustments or low-angle tasks. This is because FlexiCamAR helps reduce physical demand and has greater flexibility. In contrast, the baseline in the follow-view mode was suitable for stable and multitasking scenarios despite its physical demands. FlexiCamAR's anchor-view mode, despite the flexible feature, required more operational adjustments, impacting user comfort and efficiency.

\subsection{Summary and Discussion}
\label{sec:summary_and_discussion}
Our study aimed to answer two research questions regarding the efficacy of FlexiCamAR compared with the baseline (\textbf{RQ1}) and the impact of different display modes (\textbf{RQ2}).

Regarding \textbf{RQ1}, the results indicated that FlexiCamAR significantly reduces physical load while improving efficiency in tasks that require frequent viewpoint adjustments. This was consistently observed in the NASA-TLX data for both tasks, where the approach demanded significantly less physical effort. This ergonomic benefit also translated to a performance advantage in the QR code-scanning task, where FlexiCamAR was significantly faster.

Regarding \textbf{RQ2}, the analysis of interaction effects revealed that the baseline is highly sensitive to display modes, whereas FlexiCamAR is more robust. Significant interaction effects in photo-taking time and NASA-TLX (Mental and Frustration) indicate that coupling the camera with the head makes coordination harder when the display is anchored. In contrast, FlexiCamAR decouples the camera from the head. The manual flexibility of the ring camera allows users to compensate for display misalignment through hand adjustments rather than head movements, maintaining consistent performance and comfort across follow-view and anchor-view.

However, our findings also reveal a trade-off in FlexiCamAR, where the same flexibility that offers ergonomic benefits also creates stability challenges. Several participants, while acknowledging the comfort of the approach, expressed concerns about holding the viewpoint steady for tasks such as photography. This suggests that while the baseline enforces stability at the cost of physical effort, FlexiCamAR prioritizes ergonomic flexibility. Practical applications could benefit from computational support, such as image stabilization, to improve camera stability.

This finding is also evident in the subjective usability ratings. The SUS results indicated that its usability was consistent across both follow-view and anchor-view display modes. In contrast, the usability of the baseline was fragile and degraded significantly in the anchor-view mode. This suggests that FlexiCamAR is a more flexible and reliable interaction approach, less dependent on a specific method of displaying its viewpoint. 

It is also worth discussing the potential to use FlexiCamAR as an extension to smartphone cameras. While a finger-worn camera could technically pair with a mobile phone to provide a flexible viewpoint, the benefits of this approach are maximized when coupled with AR glasses. In a smartphone-based setup, users must keep the phone screen within their visual field to monitor the camera feed, which imposes additional constraints on head and body posture. In contrast, AR glasses fundamentally decouple the display from the input device. This feature ensures that the camera feed remains consistently accessible within the user's line of sight, regardless of their posture, thereby fully preserving the ergonomic freedom and convenience offered by the flexible viewpoint.

These findings motivate us to explore scenarios in which FlexiCamAR is not just an improvement but an enabling factor. We discuss some unique and potential use cases next in~\cref{sec:Unique Use Cases and Scenarios}.

\begin{figure*}[tb]
  \centering
  \includegraphics[width=\textwidth]{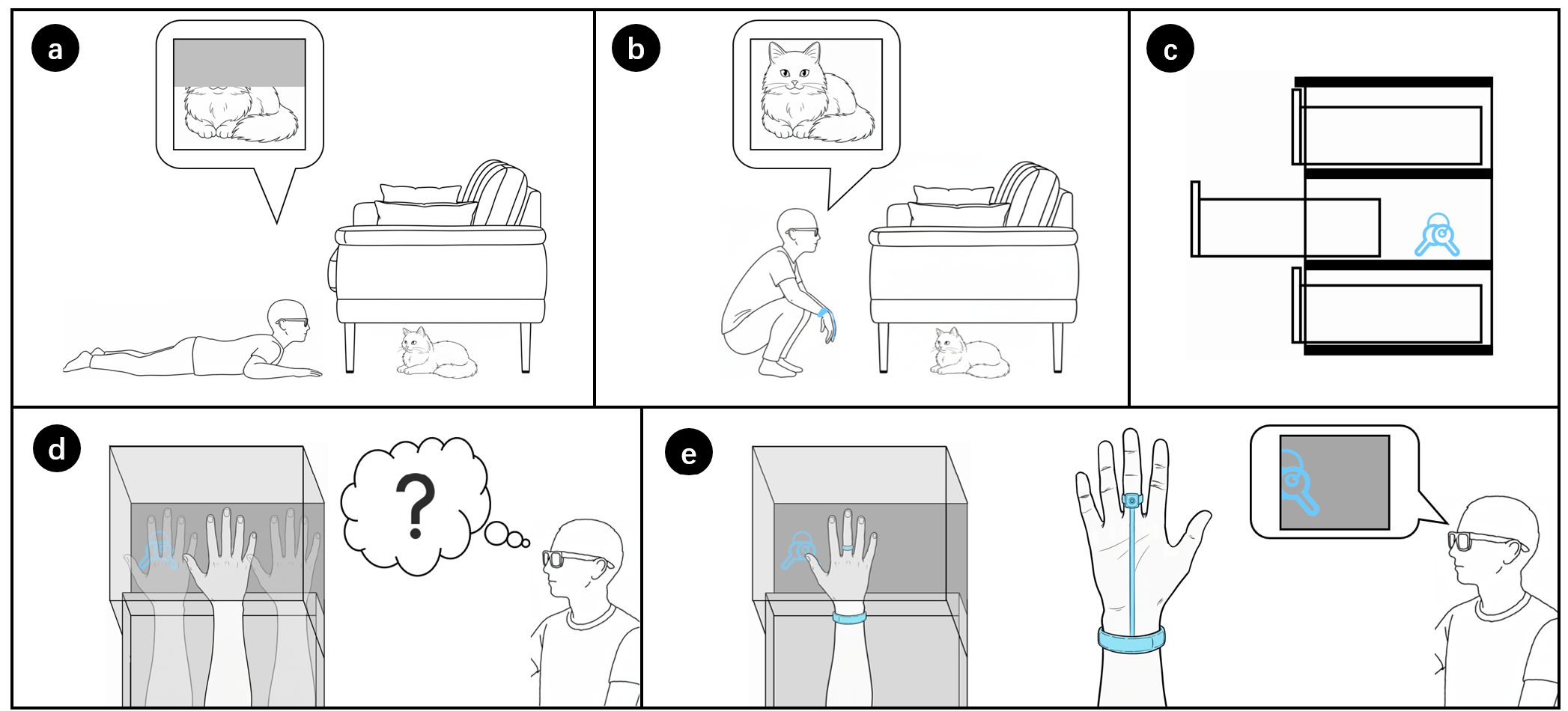}
  \caption{Some possible use cases of FlexiCamAR. (a,b) showcases capturing low-angle perspectives: (a) With the baseline, the user must adopt a physically strenuous posture, often resulting in an obstructed view. (b) FlexiCamAR enables the user to capture a complete, unobstructed view while maintaining a comfortable posture. (c-e) demonstrates inspecting a confined space: (c) A key is lodged in the narrow gap behind a drawer. (d) The baseline interaction forces the user to rely on a blind, tactile search. (e) FlexiCamAR provides a direct visual feed of the space, enabling rapid and precise, visually-guided retrieval.}
  \label{fig:SpecificScenario}
\end{figure*}

\section{Additional Potential Use Cases and Scenarios}
\label{sec:Unique Use Cases and Scenarios}
To fully explore the potential of FlexiCamAR, our study utilized a wired, high-fidelity prototype capable of streaming 1080p video at 30 fps. We acknowledge that current battery and wireless technologies do not yet support this level of performance in a compact, wireless ring form factor, with state-of-the-art systems like IRIS~\cite{kim2024} being limited to low-resolution, low-framerate grayscale streaming. Therefore, this section explores several use cases not as a demonstration of a market-ready product, but as a forward-looking exploration of the rich interaction possibilities that FlexiCamAR could unlock once these technological limitations are overcome. These scenarios serve to illustrate the fundamental ergonomic and functional value of the approach itself.

\subsection{Low-Angle and Ground-Level Perspectives}
\label{sec:usecase_low_angle}
The fixed first-person perspective in AR becomes challenging when the user needs to capture ground-level views. For example, taking an eye-level photo of a pet under a sofa requires the user to adopt strenuous postures, such as kneeling, crouching, or even lying on the floor (see~\cref{fig:SpecificScenario}.a,b). These strenuous postures can be disruptive and often lead to an obstructed view. FlexiCamAR solves this problem by separating the camera's viewpoint from the user's head. This allows the user to capture a clear, complete view from a comfortable position. This example also shows the reduction in physical load found in our user study and suggests new possibilities for everyday AR photography.

\subsection{Inspecting Confined or Obstructed Spaces}
\label{sec:usecase_confined_spaces}
FlexiCamAR is also helpful for tasks that involve looking into tight spaces. These are places where the user's head and the built-in AR camera cannot fit. A practical example is retrieving a small object that has fallen into the narrow, visually obstructed gap behind a drawer, preventing it from closing. In this situation, the baseline is rendered ineffective. The user has to search for the object by touch alone. This process of repeatedly feeling around without seeing is often inefficient and frustrating. FlexiCamAR fundamentally changes this task by augmenting the user's hand with a visual sensor. This approach allows the user to view the tight space with the camera. Instead of searching by touch, the user can now see the object and retrieve it accurately (see~\cref{fig:SpecificScenario}.c-e). This example demonstrates that FlexiCamAR can be a valuable tool. It does more than just improve comfort; it can also change how users understand and solve physical problems around them.

\section{Limitations and Future Work}
\label{sec:Limitations}
While our findings highlight the potential of FlexiCamAR, we acknowledge several limitations in our current investigation. These limitations not only define the boundaries of this work but also offer possible avenues for future research.

We intentionally focused our evaluation on specific tasks. For this paper, we chose common activities such as taking photos and scanning QR codes. However, these tasks are just a few examples of many potential everyday applications. Our results show that the approach is helpful in these situations. In the future, we plan to explore other tasks, such as video conferencing, environmental scanning, or more complex photography using AR HMDs, which would benefit from having a secondary camera view. 

The prototype we built for the study also had limitations. It had to be connected via a wire, because current battery technology is not yet compact enough or long enough to support high-quality wireless video streaming~\cite{kim2024}. This wired connection meant our study could not fully represent real-world conditions. Our future work also involves exploring FlexiCamAR to include more complex scenarios, such as walking in public spaces. This exploration will allow us to gain further insights into the integration of our approach in people's daily use. This would require overcoming some technological barriers and longer-term observations.  

Our study also brought up important social and ethical issues with the FlexiCamAR approach. With this technology, the camera's view is separate from the user's line of sight. This creates the potential for the device to be misused for taking photos unnoticed by others, a concern our participants raised. This is not just a challenge for a ring-shaped device, but a broader social responsibility that designers of any similar system must consider. Addressing these social and ethical issues will be crucial for people to accept this type of technology for everyday use.

Our work confirmed the main benefits of FlexiCamAR and gave us a better understanding of its features. This approach opens the door to rich interaction possibilities. To support other interactions, it can be combined with other sensors and interaction methods. For example, computer vision software could analyze the video feed from the finger-worn camera. This would allow the system to do more than just provide a view. It could enable new functions, such as interacting with objects, identifying materials, or recognizing small gestures. These additions could transform FlexiCamAR from a basic viewing tool into a versatile AR interaction device. This would support more advanced ``see and manipulate" actions and represents another direction we plan to pursue in the future.

\section{Conclusion}
\label{sec:Conclusion}
Front-facing cameras are the current standard for AR glasses. However, our work challenges this one-size-fits-all design by showing the ergonomic and functional limits of a fixed first-person view for many everyday tasks. In response, we developed and tested FlexiCamAR, a new approach that gives users a flexible second camera viewpoint separate from their head. Our user study showed that this approach has clear value. It can significantly reduce the physical effort needed for common tasks. We also discussed FlexiCamAR's new capabilities in potential use cases and scenarios, such as inspecting tight spaces, which are difficult with a fixed first-person view camera. This expands what is possible with AR. Ultimately, our contribution is more than just a new technique; it is a new way of thinking about camera design for AR glasses. We believe that the future of camera interaction on AR glasses is not a single fixed view. Instead, it should be a flexible combination of views that adapts to the user's task and situation. FlexiCamAR is an important first step in this direction and opens the door to richer, more unique interactions with AR glasses.

\bibliographystyle{abbrv-doi}

\bibliography{Reference}
\end{document}